\documentclass[aps,prb,a4paper,twocolumn,10pt,floatfix,showpacs,showkeys,amsmath,amssymb,nobibnotes,unsortedaddress,superscriptaddress]{revtex4-2}

\setcitestyle{super}
\usepackage{graphicx}
\usepackage[export]{adjustbox}
\usepackage[dvipsnames]{xcolor}
\usepackage[utf8]{inputenc}
\usepackage[T1]{fontenc}
\usepackage{amsmath}
\usepackage{xr}
\usepackage{hyperref}
\usepackage{upgreek}
\usepackage[percent]{overpic}
\usepackage{xcolor}
\usepackage{physics}
\usepackage{verbatim}
\usepackage{csquotes}
\usepackage[explicit]{titlesec}
\usepackage{lipsum}
\usepackage{xspace}
\usepackage{lineno}
\usepackage{appendix}

\graphicspath{{figures/}}
\renewcommand{\figurename}{Figure}
\renewcommand{\thefigure}{\arabic{figure}}
\renewcommand{\tablename}{Table}
\renewcommand{\thetable}{\arabic{table}}
\renewcommand{\thesection}{\arabic{section}}

\titleformat{\section}{\large\bfseries\filcenter}{\thesection.}{1em}{#1}

\newcommand{\Voc}{V_\mathrm{oc}}

\newcommand{\FF}{F\!F}

\newcommand{\Rtr}{R_\mathrm{tr}}

\newcommand{\nid}{n_\mathrm{id}}

\newcommand{\Eu}{E_{U}}

\newcommand{\mueff}{\mu_\mathrm{eff}}
\newcommand{\mun}{\mu_\mathrm{n}}
\newcommand{\mup}{\mu_\mathrm{p}}
\newcommand{\sigmaeff}{\sigma_\mathrm{eff}}
\newcommand{\sigman}{\sigma_\mathrm{n}}
\newcommand{\sigmap}{\sigma_\mathrm{p}}

\newcommand{\kT}{k_B T}

\begin{document}

\title{The contribution of electron and hole conductivity to the transport loss in organic solar cells}

\author{Chen Wang}
\affiliation{Institut für Physik, Technische Universität Chemnitz, 09126 Chemnitz, Germany}

\author{Toni Seiler}
\affiliation{Institut für Physik, Technische Universität Chemnitz, 09126 Chemnitz, Germany}

\author{Doyoung Sun}
\affiliation{Institute of Physics and Astronomy, University of Potsdam, Karl-Liebknecht-Str.24-25, D-14476 Potsdam-Golm, Germany}
\affiliation{Heterostructure Semiconductor Physics, Paul Drude Institute for Solid State Electronics, Hausvogteiplatz 5-7, 10117 Berlin, Germany}

\author{Safa Shoaee}
\affiliation{Institute of Physics and Astronomy, University of Potsdam, Karl-Liebknecht-Str.24-25, D-14476 Potsdam-Golm, Germany}
\affiliation{Heterostructure Semiconductor Physics, Paul Drude Institute for Solid State Electronics, Hausvogteiplatz 5-7, 10117 Berlin, Germany}

\author{Maria Saladina}
\affiliation{Institut für Physik, Technische Universität Chemnitz, 09126 Chemnitz, Germany}

\author{Carsten Deibel}
\email[Corresponding author. Email: ]{deibel@physik.tu-chemnitz.de}
\affiliation{Institut für Physik, Technische Universität Chemnitz, 09126 Chemnitz, Germany}

\begin{abstract}

The effective conductivity determines the reciprocal of the transport resistance, the dominant loss of fill factor in organic solar cells. We experimentally determine the dependence of effective conductivity on its electron and hole contributions. Using PM6:Y12 blends with tunable morphological and energetic disorder, we show that the effective conductivity follows a harmonic mean of electron and hole conductivities even across nearly three orders of magnitude in conductivity imbalance. We also validate the method for directly extracting effective conductivity from current--voltage measurements, eliminating the need to rely on indirect mobility and charge carrier density-based proxies. Our findings challenge the widespread use of geometric mean approximations and offer a more accurate framework for analysing and modelling transport in disordered organic semiconductors.

\end{abstract}


\maketitle

\section{Introduction}

Organic semiconductors have been widely employed in light-emitting diodes, field-effect transistors, and photovoltaic devices over the past decades.\cite{nakanotani2014high,muccini2006bright,leo2016organic} Unlike their inorganic crystalline counterparts, charge transport in these materials is dominated by localised hopping due to poor electron wave function overlap, rather than extended band transport. This results in intrinsically low electrical conductivity. Poor conductivity contributes to charge transport losses and limits efficient charge carrier extraction or injection. In organic solar cells (OSCs), inefficient charge transport leads to collection losses and manifests as transport resistance, which acts as an internal series resistance and substantially reduces the fill factor ($\FF$).\cite{schiefer2014determination,wurfel2015impact,wopke_traps_2022,wang2025transport} Therefore, understanding which charge carrier type limits the effective conductivity -- and thus the $\FF$ -- is essential for improving the device performance.

Beyond charge extraction, the yield of all key processes in OSCs -- photogeneration, nongeminate recombination, and charge transport -- has to do with how fast the excitons or charge carriers move. Photogeneration is efficient if the photogenerated excitons diffuse quickly to reach the donor--acceptor interface within their lifetime, and the resulting charge transfer exciton can dissociate more easily if at least one charge carrier type is fast.\cite{Veldman2008} For nongeminate recombination, Langevin's concept of encounter-limited recombination\cite{Langevin1905} depends strongly on the electron and hole mobilities $\mun$ and $\mup$, respectively. 
Heiber et al.\cite{heiber2015encounter} showed by kinetic Monte Carlo simulations that the nongeminate recombination rate is given by the classical Langevin rate in homogeneous materials or blends with very small domain size (charge carrier lifetime $\tau \propto 1/(\mun + \mup)$). However, as electrons and holes are usually moving only within their respective material phase, the electron in the acceptor and the hole in the donor, they can only meet and recombine at the donor--acceptor interface. For typical material domain sizes of 10 to 40 nm, the charge carrier lifetime was found to depend on the geometric mean of electron and hole mobilities ($\tau \propto 1/(\mun \mup)^{1/2}$), whereas only for large domain sizes does the harmonic mean (or its extreme case, the minimum mobility\cite{Koster2006}) determine the lifetime ($\tau \propto 1/(1/\mun + 1/\mup)^{-1}$). 

For recombination alone, imbalanced mobilities would lead to the lowest recombination rates. However, a solar cell can only be good if the charge carriers are eventually extracted into the external circuit to drive a load. Therefore, a common figure-of-merit for good solar cells with high $\FF$ is a high mobility--lifetime combination: a high $\mu \tau$-product. Hecht already used it to describe current--voltage characteristics in 1932.\cite{Hecht1932} The concept was later adapted to disordered materials,\cite{Crandall1982} although for OSCs the assumption that the lifetime is constant\cite{Street2010} is not correct: it is a strong function of the carrier concentration.\cite{Deibel2010,Stolterfoht2015}

If mobility\cite{Tessler2004,Bartelt2015} and lifetime are both critical for efficient charge collection yielding a high $\FF$,\cite{kaienburg2016extracting} yet $\tau \propto 1/\mu$ as outlined above, how can the $\mu\tau$-product be increased? Since nongeminate recombination is a pair-wise process requiring electrons and holes to find each other, whereas charge transport is an individual process ending with their extraction at opposite electrodes, it becomes clear that the impact of electrons and holes on $\mu$ and on $\tau$ is not the same. It has long been recognised that either balanced or sufficiently high charge carrier mobilities are required to minimise collection losses,\cite{Tessler2006,Kotlarski2011,Bartelt2015} yet typical mobility values are not high enough by more than an order of magnitude in state-of-the-art OSCs. Consequently, the $\FF$ is mainly limited by transport resistance, which scales inversely with the \emph{effective} conductivity of the active layer\cite{neher2016new,saladina_transport_2025,wang2025transport} -- and arises even in the absence of space-charge limitations.\cite{mihailetchi_space-charge_2005}

This raises a key question: \emph{how} do electron and hole mobilities contribute to the effective mobility in the $\mu\tau$-product, and more generally, how do their conductivities interact to yield an effective conductivity?
Understanding this dependence is essential for identifying charge transport bottlenecks, i.e., for determining which charge carrier type -- and thus which material, donor or acceptor -- limits charge collection.

Early analytic descriptions, for planar heterojunction OSCs,\cite{Cheyns2008} p-i-n junctions,\cite{Muller2013} and bulk heterojunctions,\cite{schiefer2014determination,wurfel2015impact} showed that the electron and hole conductivities contribute to transport resistance through their harmonic mean. For equal electron and hole currents, $j_p = j_n = j/2$, the transport resistance across the active layer thickness $L$ was given by 
$$\frac{\Rtr}{L} = \frac{1}{\sigmap} + \frac{1}{\sigman}.$$
This implies that the lower conductivity strongly dominates the transport resistance. For very imbalanced values, the larger one becomes negligible, and the lower conductivity determines the transport-related $\FF$ loss. 

Discrepancies arose in literature a decade ago, with the goal to express the transport resistance in terms of mobilities rather than conductivities. 
Schiefer et al.\cite{schiefer2014determination} expressed 
$$\frac{\Rtr}{L} = \frac{1}{2e}\left( \frac{1}{p \mup} + \frac{1}{n \mun} \right),$$
showing that the harmonic mean of conductivities translates into the harmonic mean of mobilities (for equal electron and hole concentrations $n$ and $p$). One year later, another relevant work was released, in which the authors expanded upon the earlier derivation:\cite{wurfel2015impact} 
Generally, charge transport is driven by quasi-Fermi level (QFL) gradients, e.g.\ for electrons, $j_n = \sigman/e \cdot \nabla E_F^n = \mun n \nabla E_F^n$. Based on the requirement of equal electron and hole currents already stated above, the authors assumed that the QFL gradients were the same, $\nabla E_F^n = \nabla E_F^p$, in turn requiring the conductivities to be equal, $\sigman = \sigmap$. We point out that this limitation corresponds to interpreting the active layer as effective medium with only one effective gap everywhere –- a simplification that does not account for the possibility of band bending between the donor and acceptor phases.\cite{Stelzl2012,McMahon2011,Karuthedath2020} The assumption of equal conductivities implies that when electron and hole concentrations are not equal, their mobilities must be imbalanced. The authors then derive an effective conductivity, $\sigmaeff = 2e \sqrt{np \mun \mup} = 2 \sqrt{\sigman \sigmap}$. Here, the geometric mean of conductivities would determine the transport resistance. Albrecht et al.\ also adopted equal QFL gradients for electrons and holes and, therefore, balanced conductivities to explain the matching currents.\cite{albrecht2014quantifying} They derived that the effective mobility is given by the harmonic mean of electron and hole mobilities. The two first papers connecting the transport resistance with $\FF$ losses in bulk heterojunction OSCs, Bartesaghi et al.\cite{Bartesaghi2015} and Neher et al.,\cite{neher2016new} used the geometric mean of mobilities to define their figures-of-merit. Here, the limitations imposed by slow charge carriers\cite{Stolterfoht2016} can be compensated by similarly weighted fast charge carriers. In contrast, in the harmonic mean of conductivities, the slower charge carrier dominates the charge collection losses. Only recently has it been proposed as a more accurate description of $\FF$ losses due to imbalanced electron--hole transport.\cite{saladina_transport_2025} However, a systematic experimental confirmation of this approach was lacking up to now.

In this Letter, we study PM6:Y12 OSCs with systematically varied donor fraction down to 1~\% to explore the relationship between effective conductivity and imbalanced electron/hole transport. We demonstrate experimentally that across nearly three orders of magnitude of transport asymmetry, the \emph{harmonic mean} best describes the dependence of effective conductivity on the individual electron and hole conductivities -- indicating that the slower hole transport in the polymer donor governs the overall charge transport. Our findings hold under variations in both spatial and energetic disorder within the photoactive layer. 
Furthermore, we validate a recently proposed method for determining effective conductivity, based on light intensity-dependent current--voltage and open-circuit voltage measurements, offering a direct and reliable approach to probe the limiting charge carrier species in low-mobility semiconductor devices.

\section{Results and Discussion}

\begin{figure}[!tb]
    \centering
    \includegraphics[width=0.95\linewidth]{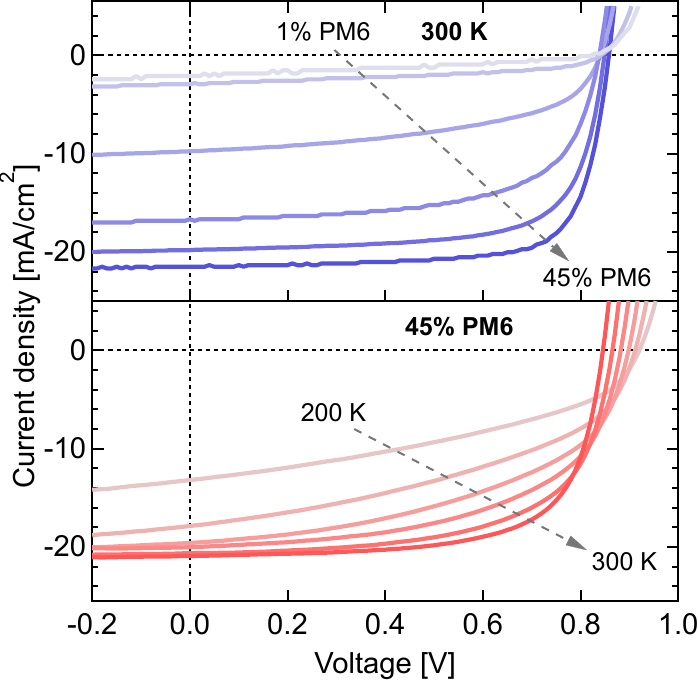}
    \caption{Representative JV curves of solar cells based on different PM6 fraction (top panel), and different measurement temperature (bottom panel) under 1~sun equivalent illumination. 1~sun corresponds to $100\,\mathrm{mW(cm)^{-2}}$ under a AM1.5G conditions.}
    \label{fig:figure01}
\end{figure}

PM6:Y12, a donor--acceptor system with reported power conversion efficiencies exceeding 16~\%,\cite{ma_achieving_2020} is selected as a model system to investigate the relationship between the individual electron and hole conductivities and the effective conductivity. To systematically probe this relation, we prepared bulk heterojunction solar cells with donor concentrations ranging from highly diluted (1~\%) to optimized (45~\%) levels. This compositional variation provides control over the formation of percolation pathways for hole transport, enabling modulation of hole conductivity over several orders of magnitude.

To further broaden the experimental scope, we performed temperature-dependent measurements, to vary the effective energetic disorder within the active layer systematically and, in turn, modify the charge carrier conductivities. In organic semiconductors, charge transport occurs through a thermally activated hopping mechanism between localized states that are energetically disordered.\cite{baranovskii_mott_2018} Accordingly, the charge carrier mobility -- and thus the conductivity -- is thermally activated and increases with temperature. By varying both the spatial morphology through blend composition and the energetic disorder through temperature, we are able to systematically investigate the dependence of conductivity across a broad range of charge transport conditions. 

The photovoltaic performance of the devices was first characterized, and representative current density--voltage (JV) curves are shown in Figure~\ref{fig:figure01}. Both the PM6 fraction and the temperature strongly influence the JV characteristics. In particular, the slope near the open-circuit voltage ($\Voc$) varies between devices. In low-mobility OSCs, this slope is primarily governed by internal transport resistance due to imperfect charge transport within the active layer rather than by constant external series resistance from electrodes and interfaces.\cite{schiefer2014determination,saladina_transport_2025,wang2025transport} A shallower slope indicates higher transport resistance and thus lower effective conductivity, resulting in a larger $\FF$ loss in the solar cells. We exclude a space-charge (SC) limited photocurrent as the origin of the loss, as such devices are predicted to have a $\FF$ of 42~\%.\cite{mihailetchi_space-charge_2005} Some of our devices show even lower values, indicating that transport resistance is the dominant $\FF$ limitation (cf.\ Figure~\ref{fig:SCL_limitation}). 

We experimentally determine the electron ($\sigman$) and hole ($\sigmap$) conductivities, along with the effective conductivity ($\sigmaeff$), and compare them using different averaging approaches, including the arithmetic, geometric, and harmonic means. 
The values of $\sigman$ and $\sigmap$ were independently measured in the dark using single-carrier devices fabricated with the same photoactive layers as the solar cells. The device configurations and corresponding JV characteristics are shown in Figure~\ref{fig:SI01}. Conductivity was determined by the linear dependence of the current density on the voltage around 0~V in these unipolar devices, representing a constant field across the active layer 
\begin{equation}
    J = \sigma_\mathrm{n,p} \cdot F = \sigma_\mathrm{n,p} \cdot \frac{V}{L} , 
\end{equation}
where $F$ is the electric field, and $L$ is the active layer thickness. On the same devices, we observed SC limited currents and determined the electron and hole mobilities accordingly.\cite{mark1962space,nicolai2011electron}

We recently presented an analytical method based on the concept of transport resistance, which enables the determination of $\sigmaeff$ from the light intensity-dependent JV and $\Voc$.\cite{saladina_transport_2025} Our approach is versatile and enables direct measurement of conductivities well below $10^{-7}$~S/cm, beyond the range accessible to conventional four‑point probe techniques.\cite{van1958method,schwarze2019molecular,yang2024carrier} Figures~\ref{fig:SI02} and \ref{fig:SI03} summarize the JV curves used in this study. For any given illumination intensity, $\sigmaeff$ is calculated from the following expression:\cite{saladina_transport_2025}
\begin{align}\label{eq:sigmaoc}
    \frac{L}{\sigmaeff}=\left.\left(\frac{dV}{dJ}-\frac{\nid\kT}{eJ_\mathrm{rec}}\right)\right|_{J=0} . 
\end{align}
Here, $n_\mathrm{id}$ is the ideality factor, extracted from the slope of the light intensity-dependent $\Voc$,\cite{tvingstedt2016temperature} and $\kT/e$ is the thermal voltage. The recombination current density $J_\mathrm{rec}$ equals the generation current density when the net current is zero. We estimate the latter from the current density measured at a reverse bias of -0.5~V. A detailed background of the method can be found in our earlier works.\cite{saladina_transport_2025,wang2025transport}

The extracted $\sigmaeff$ values obtained via this method exhibit a clear dependence on light intensity (cf.\ Figure~\ref{fig:SI04}), primarily caused by changes in the charge carrier density $n$. However, an additional contribution may come from charge carrier mobility. As charge transport in disordered semiconductors occurs through thermally activated hopping, the mobility -- and thus effective conductivity -- is sensitive to the density and energetic depth of trap states and how it determines the concentration of mobile charge carriers. These factors, in turn, are governed by the shape of the density of states (DOS). We have previously shown that the DOS in state-of-the-art OSCs is better described as a combination of a Gaussian and a power-law distribution.\cite{saladina2023power} In contrast to an exponential DOS, where the characteristic energy (Urbach energy, $\Eu$) remains constant, a power-law DOS exhibits an energy-dependent broadening: the $\Eu$ increases towards the midgap.

\begin{figure}[!tb]
    \centering
    \includegraphics[width=0.95\linewidth]{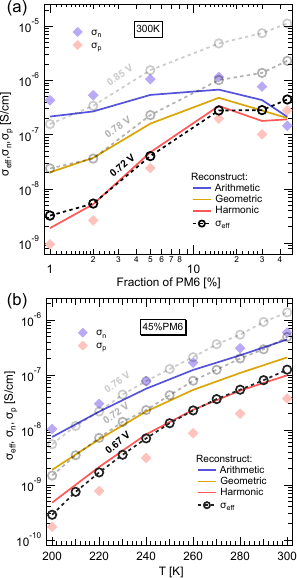}
    \caption{(a) Comparison of plots of experimental $\sigmaeff$ (dashed lines with open circles), separated $\sigman$ and $\sigmap$ (diamonds), and reconstructed effective conductivities (solid lines) based on different means as a function of PM6 content under 300~K. The different dashed lines represent $\sigmaeff$ extracted at different $\Voc$ values. (b) reconstruction of $\sigmaeff$ based on $45\%$ PM6 under different temperatures.}
    \label{fig:sigma_reconstruct}
\end{figure}

The $\sigman$ and $\sigmap$, along with the $\sigmaeff$ determined using the methods described above, are presented in Figures~\ref{fig:sigma_reconstruct}(a) and (b). Having established reliable approaches for extracting these values, we now examine how they vary across device composition and temperature, respectively. 
The individual $\sigman$ and $\sigmap$ are represented by diamond symbols. Reducing the PM6 content alters the morphology of the photoactive layer, significantly impacting the connectivity of PM6 domains and thereby hindering hole transport. In contrast, the Y12 domains that govern electron transport become more abundant but remain relatively unaffected. As a result, $\sigmap$ decreases by nearly three orders of magnitude going from a donor fraction of 45~\% to 1~\%, while $\sigman$ stays largely constant -- creating a transition from balanced to strongly unbalanced charge transport across the devices. In Figure~\ref{fig:sigma_reconstruct}(b), temperature variation introduces changes in the effective energetic disorder that charge carriers experience, affecting both $\sigman$ and $\sigmap$. Over the measured temperature range, both conductivities change by roughly two orders of magnitude, consistent with the thermally activated hopping transport characteristic of disordered organic semiconductors.

We now turn our attention to the effective conductivity, extracted for the same set of compositional variations and temperatures. With this complete dataset ($\sigman$, $\sigmap$, and $\sigmaeff$), our goal is to evaluate how effective conductivity depends on the underlying electron and hole conductivities. This requires ensuring that the different measurement conditions of the $\sigmaeff$ -- measured in solar cells under illumination -- and the $\sigman$ and $\sigmap$ -- measured from electron-only and hole-only single-carrier devices, respectively, in the dark -- are accounted for. Single-carrier devices employ electrodes with nearly identical work functions, resulting in minimal built-in field and an energy landscape near flat-band conditions. In these devices, the charge carrier density is determined primarily by injection from the electrodes rather than by illumination that generates a QFL splitting and governs the charge carrier density in solar cells. We find that the most comparable conditions are close to flat band conditions and therefore compare the $\sigman$ and $\sigmap$ at 0~V for the single-carrier devices with $\sigmaeff$ at certain $\Voc$ values for the solar cells. This has the additional advantage that we probe at similar energetic positions relative to the transport levels, because the photovoltaic bandgap of all devices is similar (cf.~Figure~\ref{fig:SI05}). By this approach of comparing the conductivities at given $\Voc$ values, we can find the conditions under which the carrier concentrations between the two device types can be compared meaningfully.

The extracted $\sigmaeff$ values at $\Voc$ from solar cells -- as represented by dashed lines with open circles -- are compared with predictions based on three averaging expressions: the arithmetic mean $\sigmaeff = (\sigman + \sigmap)/2$, the geometric mean $\sigmaeff = \sqrt{\sigman \sigmap}$, and the harmonic mean $\sigmaeff = 2 / (\sigman^{-1} + \sigmap^{-1})$. The $\sigmaeff$ -- reconstructed from the individual $\sigman$ and $\sigmap$ values using these expressions -- are shown as solid lines in Figures~\ref{fig:sigma_reconstruct}(a) and (b). We find that extracting $\sigmaeff$ at different $\Voc$ values primarily leads to a vertical offset in magnitude, while the overall trend as a function of PM6 fraction or temperature remains consistent. 
Notably, this trend closely follows the behaviour predicted by the harmonic mean. By selecting the conductivity values at cross sections of the $\Voc$ of 0.72~V for the compositional variation and 0.67~V for the temperature -- we observe quantitative agreement between the measured $\sigmaeff$ and the harmonic-mean of the separately measured $\sigman$ and $\sigmap$, respectively. This consistency across two independent experimental axes strongly supports the harmonic mean as the most appropriate descriptor for the effective conductivity in these devices. Moreover, the agreement between reconstructed and measured $\sigmaeff$ values validates the reliability of the transport resistance-based method, confirming that it provides a direct way to access effective conductivity in low-mobility OSCs. 

\begin{figure}[!tb]
    \centering
    \includegraphics[width=0.95\linewidth]{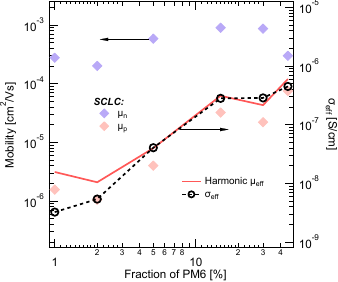}
    \caption{Comparison between electron and hole mobility and effective $\sigmaeff$, harmonic $\mueff$ calculated from $\mun$ and $\mup$ determined by SC limited current measurements.}
    \label{fig:mobility_reconstruct}
\end{figure}

While our study focuses on conductivity, it is closely linked to the effective charge carrier mobility through the relation $\sigmaeff = en\mueff$. In organic semiconductors, mobility is often used as a proxy for describing charge transport, primarily because direct measurement of conductivity is difficult in low-mobility materials. As a result, mobility has become the dominant transport parameter used in both experiments and simulations. However, the assumed relationship between individual charge carrier mobilities and effective transport is rarely verified experimentally. To investigate this connection, we reconstruct the effective mobility ($\mueff$) -- using the individual $\mun$ and $\mup$, from SC-limited current measurements -- and compare it to the experimentally determined $\sigmaeff$. Importantly, since conductivity depends on both mobility and charge carrier density, observing a harmonic mean relationship for $\sigmaeff$ does not necessarily imply that the same relation holds for $\mueff$. The electron and hole density can differ due to a charge transport imbalance. Such a case will not only occur for the equal electron and hole conductivities often proposed in literature when combined with imbalanced mobilities, but -- as we confirmed -- for unequal conductivities. Therefore, we believe it to be essential to examine whether the harmonic mean not only describes the effective conductivity, but also accurately represents the mobility dependence under the same conditions.

The models by Mott--Gurney\cite{mott1948electronic} and Mark--Helfrich\cite{mark1962space} were applied to electron-only and hole-only JV curves to extract $\mun$ and $\mup$, respectively. The evaluation is described in the Supplemental Material, and the results are shown in Figure~\ref{fig:SI06}. 
The resulting $\mun$ and $\mup$ values as functions of PM6 content are plotted in Figure~\ref{fig:mobility_reconstruct}. As with the conductivity results, increasing the PM6 fraction has little effect on $\mun$, while $\mup$ increases significantly -- reflecting enhanced hole transport in more PM6-rich blends. These mobility results were cross-validated using resistance-dependent photovoltage (RPV) measurements, which provide direct access to transit times of faster and slower carriers to the electrodes.\cite{philippa2014impact} Both techniques yield consistent trends and comparable absolute values for the electron (fast) and hole (slow) mobilities, as shown in Figure~\ref{fig:SI07}. Importantly, we find that the slower hole mobility $\mup$ predominantly determines $\sigmaeff$. When reconstructing $\mueff$ using the harmonic mean of $\mun$ and $\mup$, the result closely matches the experimental $\sigmaeff$, confirming that the dependence of effective mobility in these systems also follows a harmonic mean.

These results have direct implications for how mobility and conductivity are interpreted in experiments and simulations. Commonly used geometric mean approximations can significantly overestimate transport properties in systems with imbalanced carrier mobilities, leading to systematic errors in extracted parameters and misleading conclusions about performance-limiting processes.

\section{Conclusion}

Unbalanced charge carrier transport is common in organic electronic devices, yet its impact is often oversimplified by applying geometric mean approximations to describe effective mobility and conductivity. We experimentally demonstrate -- for the first time to our knowledge -- that effective conductivity instead follows a harmonic mean dependence on the individual electron and hole conductivities, confirming that the slower charge carrier governs the overall charge transport and collection. This relation holds across both morphological and energetic disorder and is supported by independent mobility measurements. We further validate a robust experimental method to directly quantify effective conductivity, based on light-intensity-dependent current--voltage and open-circuit voltage measurements, providing a reliable alternative to mobility-based estimations. Our findings highlight that assuming a geometric mean can underestimate transport losses and introduce systematic errors in extracted parameters such as recombination rates or charge carrier densities. Going forward, incorporating physically accurate transport models that account for charge carrier asymmetry will be essential for accurate interpretation of loss mechanisms and improving the performance of next-generation organic photovoltaic devices.

\section*{Acknowledgements}
The authors are grateful to Professor Dieter Neher, University of Potsdam, for his valuable comments on the Letter. We thank the Deutsche Forschungsgemeinschaft (DFG) for funding this work (Research Unit FOR 5387 POPULAR, project no. 461909888).

\bibliographystyle{apsrev4-2}
\bibliography{references}

\clearpage
\onecolumngrid 
\graphicspath{{figures/}}
\renewcommand{\thepage}{S\arabic{page}}  
\renewcommand{\thesection}{S\arabic{section}}   
\renewcommand{\thetable}{S\arabic{table}}   
\renewcommand{\thefigure}{S\arabic{figure}}
\renewcommand{\theequation}{S\arabic{equation}}
\renewcommand{\figurename}{Figure}
\renewcommand{\tablename}{Table}
\setlength{\parskip}{0.3cm}
\setlength{\parindent}{0pt}
\setcounter{page}{1}
\setcounter{figure}{0}
\setcounter{table}{0}
\setcounter{equation}{0}
\setcounter{section}{0}

%\begin{document}

\begin{center}
    \large
    Supplemental Material
    \vspace{0.2cm}
\end{center}

\maketitle

\section{Experimental methods}\label{sec:S1}

\subsection{Materials} 

PM6, Y12, and PEDOT:F were purchased from Brilliant Matters Inc (Canada). ZnO nanoparticles (N10) were purchased from Avantama AG. PDINO was purchased from ONE-material. Chloroform (CF) and methanol solvents were all purchased from Merck Group. PEDOT:PSS (4083) was purchased from Heraeus Deutschland GmbH \& Co. KG. All chemicals and solvents were used as received without further purification.

\subsection{Device fabrication}

\textbf{Solar cells:} Inverted solar cells with the configuration of ITO/ZnO/PM6:Y12/PEDOT:F/Ag were fabricated. The patterned ITO substrates were pre-cleaned by ultrasonic baths in diluted Hellmanex, deionized water, acetone, and 2-propanol sequentially, and then were dried with a flowing nitrogen stream. ZnO nanoparticles dispersion ($2.5\:wt\%$ in 2-propanol) was spin-coated onto the ITO substrate at 2500~rpm for 60~s, followed by thermal treatment at $200 \,^\circ\mathrm{C}$ for 30~mins. The substrates were transferred to a nitrogen-filled glovebox for the deposition of PM6:Y12 and PEDOT:F layers. The separated PM6 (12~mg/mL in CF) and Y12 (12~mg/mL in CF) solutions were mixed by volume ratio to obtain solutions with different PM6 contents. Warm PM6:Y12 solutions ($45 \,^\circ\mathrm{C}$) were spin coated onto ITO/ZnO. 1000~rpm for 60~s, 1000~rpm for 60~s, 1000~rpm for 60~s, 1400~rpm for 50~s, 1900~rpm for 40~s, and 3000~rpm for 30~s were used in $1\%$, $2\%$, $5\%$, $15\%$, $30\%$, and $45\%$ PM6 conditions, respectively, to obtain similar active layer thicknesses of approximately 60~nm. The active layers were thermal annealed at $100 \,^\circ\mathrm{C}$ for 10~mins. PEDOT:F solution was spin coated onto the active layers dynamically at 3000~rpm for 50~s, then followed by thermal treatment at $100 \,^\circ\mathrm{C}$ for 5~mins. The solar cells were finally completed by thermal evaporation of 150~nm silver (Ag) in a vacuum chamber (vacuum degree $\approx 2\times10^{-6}$ Torr) with a shadow mask defining the active area of 0.04~$cm^2$.

\textbf{Single-carrier devices:} For electron-only devices, the configuration is ITO/ZnO/PM6:Y12/PDINO/Al. For hole-only devices, the configuration is ITO/PEDOT:PSS/PM6:Y12/PEDOT:F/Ag. For PDINO layer, the PDINO solution was prepared with 1~mg/mL in methanol. PDINO was spin coated onto the active layers dynamically at 4000~rpm for 15~s, then followed by thermal treatment at $60 \,^\circ\mathrm{C}$ for 5~seconds. For PEDOT:PSS layer, clean ITO substrates were treated using oxygen plasma for 5~mins. PEDOT:PSS solution was spin coated (3000~rpm for 50~s) and annealed at $140 \,^\circ\mathrm{C}$ for 15~min. The other layers are prepared in the same way as in solar cells.

\subsection{Device characterization}

\textbf{Current-voltage IV measurements:} The IV curves of solar cells and single-carrier devices based on different PM6 fractions were recorded using a Keithley 236 SMU in a nitrogen-filled glovebox. Both dark and light (under 1 sun) IV were recorded. AM 1.5G illumination was provided by a Wavelabs LS-2 solar simulator. No aperture was used. The illumination time for each measurement was 1.2 seconds to avoid device temperature rise.   

\textbf{Light intensity-dependent IV measurement:} The sample was mounted in a Linkam LTS420 cryostat and contacted using needle probes. A continuous wave laser (Omicron LDM A350) was used to excite the sample, operating at a wavelength of 515~nm. Motor-driven neutral-density filter wheels by Thorlabs and Standa were used to modulate the illumination intensity, and a silicon photodiode was used to monitor the illumination intensity. IV curves were measured using a Keithley 2634b SMU. Illumination equivalence between the solar simulator and laser was determined by matching $V_\mathrm{oc}$. This cryostat also enables the temperature-dependent IV measurement for solar cells and single-carrier devices. Low temperatures were ensured via a constant flow of liquid nitrogen using a Linkam Scientific LNP96-S liquid nitrogen pump and Linkam Scientific T96-S temperature controller.

\textbf{sensEQE:} The sensitive EQE measurements were carried out with a setup consisting of a 150W quartz-tungsten-halogen lamp, a 300~mm double monochromater (Quantum Design Europe MSHD-300), a mechanical chopper wheel (Thorlabs MC2000B-EC) and a Zurich Instruments MFLI lock-in amplifier. The light was chopped at a frequency of 223~Hz and a series of long pass filters (OD4) with increasing cut-on wavelength were used in order to remove influences of stray light on the measurement. After passing the Monochromator the light is passed through a liquid waveguide (Newport 77638) and focused on the sample using a lense. A small amount of light is diverted to a reference sandwich diode (Hamamatsu K1718-B) to monitor the light intensity illuminating the sample. The sample is placed inside a Microscope stage (Linkam LTS420) and contacted using Needle probes. 

\textbf{RPV:} Resistance-dependent photovoltage (RPV) measurements were carried out by exciting the solar cell with short laser pulses ($\sim 6$~ns at 532~nm), generated by a diode-pumped, Q-switched Nd:YAG laser (NT242, EKSPLA) operating at a repetition rate of 500~Hz, under open-circuit conditions. The transient photovoltage signal was recorded using an oscilloscope (Agilent DSO9104H) with an internal input impedance of 1~$\mathrm{M\Omega}$. To enhance the signal-to-noise ratio, a broadband voltage amplifier (HVA-200M-40-F, FEMTO) was employed. The effective charge carrier mobility was extracted from the initial photovoltage rise profiles by analyzing their time dependence.

\clearpage
\section{Appendix figures}\label{sec:S2}

\subsection{Space-charge Limit}

\begin{figure}[h]
    \centering
    \includegraphics[width=0.45\linewidth]{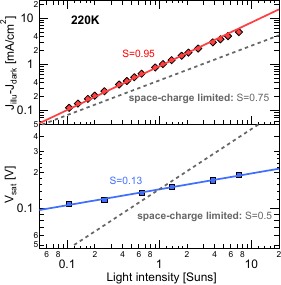}
    \caption{Dependence of $J^*$ and $V_\mathrm{sat}$ on the light intensity based on $2\%$ PM6 at 220~K, SC limited model lines are presented for the comparison.}
    \label{fig:SCL_limitation}
\end{figure}

\begin{figure*}[h]
    \centering
    \includegraphics[width=0.85\linewidth]{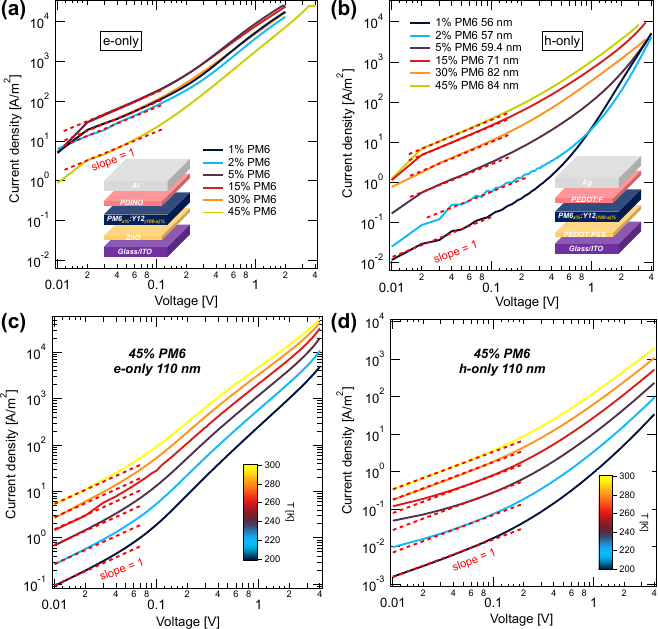}
    \caption{Representative JV curves of e-only (a) and h-only (b) devices based on different PM6 fractions. The inset is the single-carrier device configuration. Representative JV curves of e-only (c) and h-only (d) devices using $45\%$ PM6-based active layer measured under different temperatures. The corresponding active layer thicknesses were presented.}
    \label{fig:SI01}
\end{figure*}

\begin{figure*}[h]
    \centering
    \includegraphics[width=0.85\linewidth]{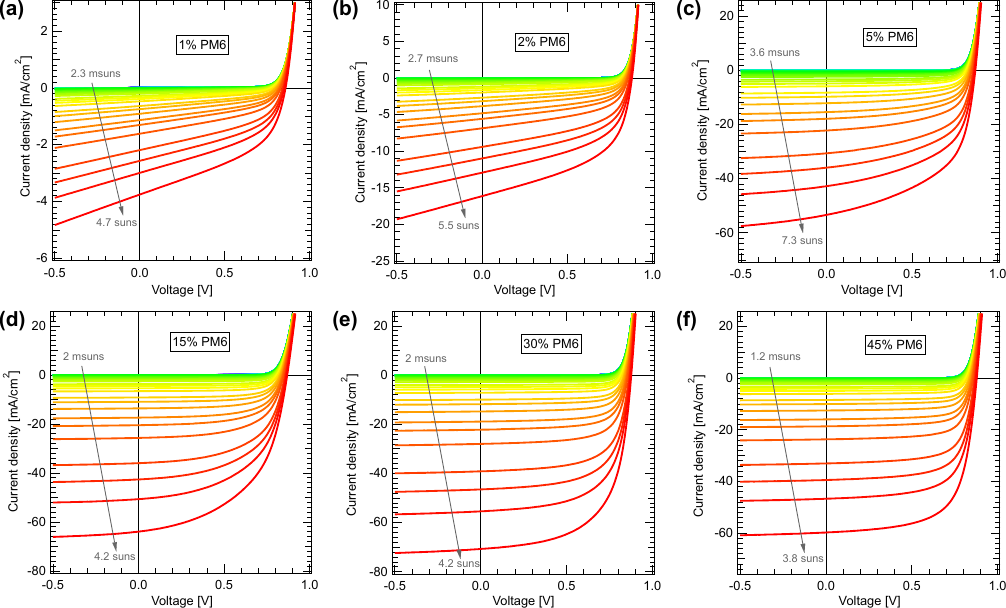}
    \caption{The (a-f) show the light intensity dependent JV curves of solar cells based on different PM6 fractions measured at 300~K, arrows indicate the light intensity range.}
    \label{fig:SI02}
\end{figure*}

\begin{figure*}[h]
    \centering
    \includegraphics[width=0.85\linewidth]{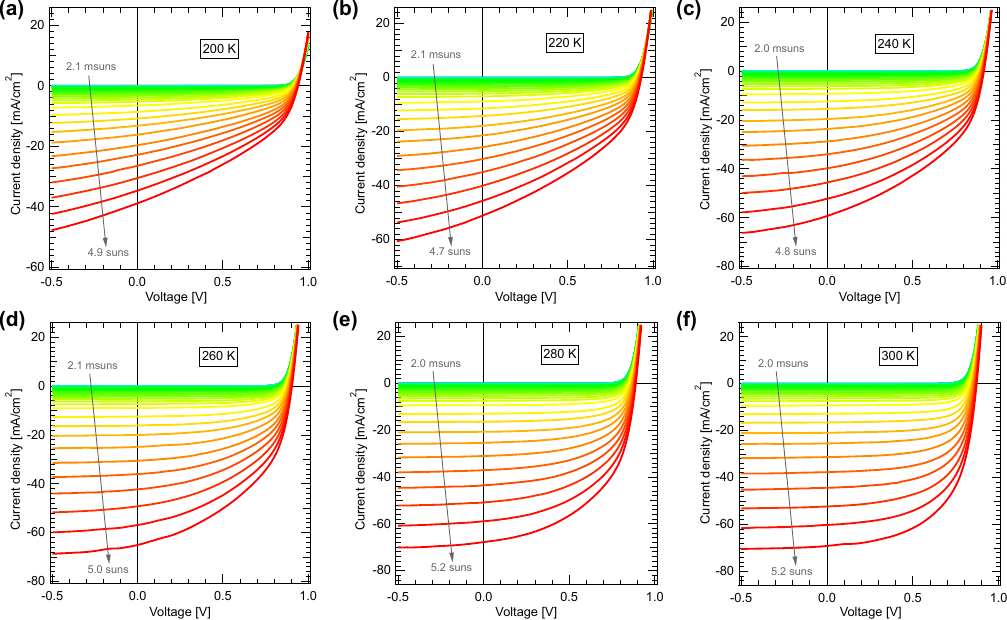}
    \caption{The (a-f) show the light intensity dependent JV curves of solar cells based on $45\%$ PM6 measured at different temperatures.}
    \label{fig:SI03}
\end{figure*}

\begin{figure*}[h]
    \centering
    \includegraphics[width=0.85\linewidth]{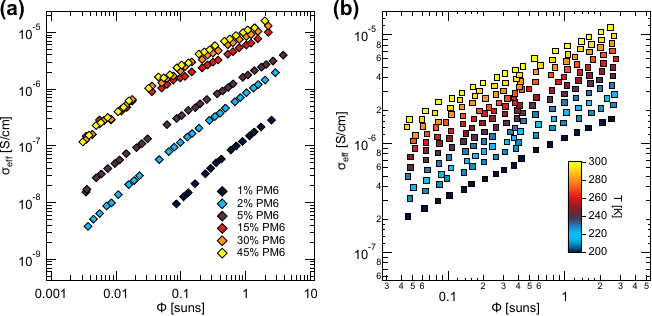}
    \caption{Light intensity dependence of $\sigmaeff$, (a) under different fraction of PM6, measured at 300~K, and (b) $45\%$ PM6 measured under different temperatures.}
    \label{fig:SI04}
\end{figure*}

\begin{figure*}[h]
    \centering
    \includegraphics[width=0.85\linewidth]{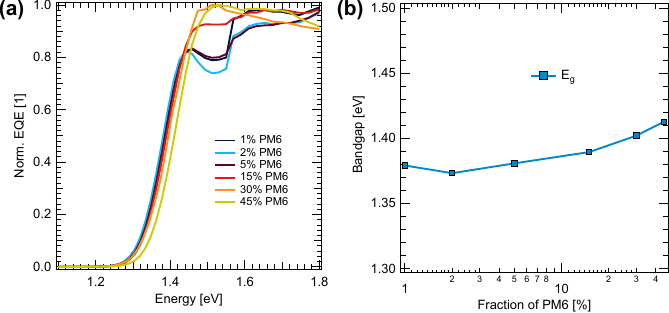}
    \caption{(a) Normalized EQE spectra of solar cells based on different PM6 fractions.  (b) $E_\mathrm{g}$ plot as a function of PM6 fraction.}
    \label{fig:SI05}
\end{figure*}

\begin{figure*}[h]
    \centering
    \includegraphics[width=0.85\linewidth]{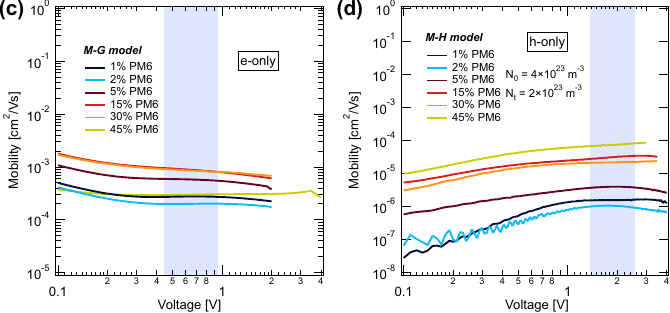}
    \caption{Electron mobility (a) and hole mobility (b) as a function of applied voltage for different PM6 fractions.}
    \label{fig:SI06}
\end{figure*}

\begin{figure*}[h]
    \centering
    \includegraphics[width=0.85\linewidth]{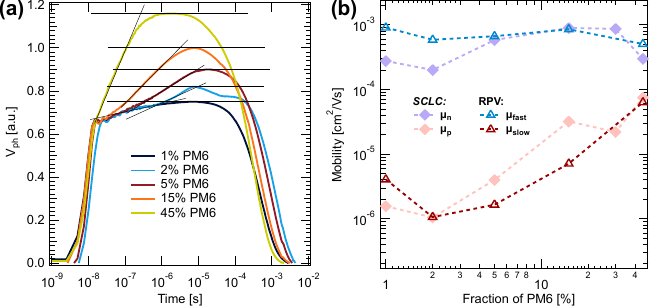}
    \caption{(a) Experimentally measured RPV transient photo-signals in solar cells based on different PM6 fractions. The tangent lines are presented here as a guide for the eye. (b) The dependence of mobility on the PM6 fraction as determined using different techniques.}
	\label{fig:SI07}
\end{figure*}

To investigate the Space-charge limitation, we also analyzed the light intensity dependence of the JV curves of 2~\% PM6-based device measured under low temperature. According to Mihailetchi et al.,\cite{mihailetchi_space-charge_2005} in the absence of SC effects, the current density $J^* = J_\mathrm{illum} - J_\mathrm{dark}$ scales linearly with light intensity. In contrast, under SC-limited conditions, $J^*$ follows a $3/4$ power-law dependence on the light intensity. Additionally, the transition voltage $V_\mathrm{sat}$ -- marking the shift from square-root dependence to saturation regime -- scales with the square-root of light intensity. Figure~\ref{fig:SCL_limitation} shows both $J^*$ and $V_\mathrm{sat}$ as a function of the light intensity. The current $J^*$ scales nearly linearly with illumination, and $V_\mathrm{sat}$ exhibits only a weak dependence, both of which significantly deviate from the SC-limited model. These results confirm that SC effects do not play a dominant role in the transport characteristics of the devices in our study.

\clearpage
\subsection{sensEQE and $E_\mathrm{g}$ determination}

We determined the energy bandgap $E_\mathrm{g}$ based on the method reported by Rau method\cite{rau_efficiency_2017}, that the bandgap energy is the mean peak energy at the absorption edge of the distribution $P(E)$:

\begin{align}
    &P(E)=\frac{dEQE}{dE}\\
    &E_\mathrm{g} = \frac{ \int_a^b E P(E)\, \mathrm{d}E}{ \int_a^b P(E)\, \mathrm{d}E}.
\end{align}
\noindent
Here, the integration limits $a$ and $b$ are chosen as the energy where $P(E)$ is equal to $50\%$ of its maximum, i.e., $P(a)=P(b)=\frac{1}{2}max[P(E)]$. It can be seen that different PM6 content shows approximately the same $E_\mathrm{g}$ $\sim 1.4$~eV.

\subsection{charge carrier mobility determination}

We calculated the charge carrier mobility as a function of applied voltage ($V_\mathrm{bi} \approx 0$ in our single-carrier devices). It can be seen that mobility shows voltage dependence in the low voltage regime, and reaches a plateau around 0.6~V for e-only devices and 2~V for h-only devices, where we extracted the field-independent $\mu_\mathrm{n}$ and $\mu_\mathrm{p}$.

The Mott--Gurney equation was used for electron-only devices:\cite{mott1948electronic}

\begin{equation}
    J=\frac{9}{8}\epsilon_{r}\epsilon_{0}\mu_\mathrm{n}\frac{V^2}{L^3}.
\end{equation}\label{eq:MG model}
\noindent
Here, $J$ is the current density, $\epsilon_{r}$ is the relative permittivity, assumed to be 3.5, $\epsilon_{0}$ is the vacuum permittivity, $\mu_\mathrm{n}$ is the electron mobility, $V$ is the applied voltage, and $L$ is the active layer thickness.

While for hole-only JV curves, rather than the quadratic dependence of the Mott--Gurney law, the slopes ($dlnJ/dlnV$) larger than 2 were observed, especially in low PM6 fractions devices. Therefore, we chose Mark--Helfrich model to treat the hole-only JV curves, in which an exponentially distributed DOS is used:\cite{mark1962space} 
\begin{equation}
    J=N_0e\mu_\mathrm{p}\left(\frac{\epsilon_0\epsilon_r}{eN_t}\right)^r\left(\frac{2r+1}{r+1}\right)^{r+1}\left(\frac{r}{r+1}\right)^r\frac{V^{r+1}}{L^{2r+1}}.
\end{equation}\label{eq:MH model}
\noindent
This is also consistent with our previous finding that the DOS in state-of-the-art OSCs is better described as a combination of a Gaussian and a power-law distribution.\cite{saladina2023power}. We used $dlnJ/dlnV$ to calculate the power of $r+1$, $N_\mathrm{0}$ and $N_\mathrm{t}$ were chosen according to the reported values.\cite{nicolai2011electron}. This allows us to solve the $\mu_\mathrm{p}$ as a function of applied voltage. Here, $J$ is the current density, $N_\mathrm{0}$ is the total density of localised states, $N_\mathrm{t}$ is the total density of trap states, $r=T_\mathrm{t}/T$, $T_\mathrm{t}$ is the characteristic temperature corresponding to the exponential DOS, $e$ is the elementary charge, $\epsilon_{r}$ is the relative permittivity, assumed to be 3.5, $\epsilon_{0}$ is the vacuum permittivity, $\mu_\mathrm{p}$ is the hole mobility, $V$ is the applied voltage, and $L$ is the active layer thickness.   

\subsection{RPV}

In the RPV measurement, the initial photovoltage rise profiles were used to extract the effective charge carrier mobility by analyzing their time dependence. Specifically, tangents were drawn to the initial quickly rising edge, the following slowly rising region, and the peak region of each voltage transient curve. The intersection points give the transit time of the faster carriers ($\tau_\mathrm{tr}^\mathrm{fast}$) and the slower carriers ($\tau_\mathrm{tr}^\mathrm{slow}$). The carrier mobility was evaluated according to the equation:

\begin{equation}
    \mu_\mathrm{fast,slow}=\frac{L^\mathrm{2}}{2V_\mathrm{bi}\tau_\mathrm{tr}^\mathrm{fast,slow}}.
\end{equation}
\noindent
Where, $\mu_\mathrm{fast,slow}$ is the fast or slow carrier mobility, $V_\mathrm{bi}$ is the built-in potential, $V_\mathrm{bi}\approx1.16\,V$, $L$ is the active layer thickness. In the $1\%$ and $2\%$ PM6 devices, the extracted slow carrier mobilities may have a large error bar, as the RPV transient was less sensitive to the small amount of charge extraction.

\bibliographystyle{apsrev4-2}

\end{document}